# The Diversity of Exoplanetary Environments and the Search for Signs of Life Beyond Earth

Sara Seager[1,2,3,*], Janusz J. Petkowski[4,5], William Bains[6,7],

[1] Department of Earth, Atmospheric and Planetary Sciences, Massachusetts Institute of Technology, 77 Massachusetts Avenue, Cambridge, MA 02139, USA
[2] Department of Physics, Massachusetts Institute of Technology, 77 Massachusetts Avenue, Cambridge, MA 02139, USA
[3] Department of Aeronautics and Astronautics, Massachusetts Institute of Technology, 77 Massachusetts Avenue, Cambridge, MA 02139, USA
[4] Faculty of Environmental Engineering, Wroclaw University of Science and Technology, 50-370 Wroclaw, Poland
[5] JJ Scientific, Mazowieckie, Warsaw 02-792, Poland.
[6] School of Physics & Astronomy, Cardiff University, 4 The Parade, Cardiff CF24 3AA, UK
[7] Rufus Scientific, Melbourn, Herts SG8 6ED, UK

*Correspondence: seager@mit.edu

## Abstract

Thousands of exoplanets orbit nearby stars, showcasing a remarkable diversity in mass, size, and orbits. With the James Webb Space Telescope now operational, we are observing exoplanet atmospheres and aiming to reach down to small, habitable-zone exoplanets in search of signs of habitability and possibly even biosignature gases. Given the scarcity of targets, it is imperative to embrace the known diversity and consider the range of exoplanets that might host life. We review how Earth life interacts with various atmospheric gases, noting that bacteria can survive in high concentrations of gases such as $H_2$, He, $CO_2$, and CO. Additionally, we consider the potential for life in alternative solvents and in cloud biospheres where rocky surfaces are excessively hot, as well as in hypothesized planetary global oceans. We highlight that life fundamentally requires metal ions for catalytic reactions, suggesting that environments without surface contact need meteoritic delivery to provide these essential elements. Despite today's observational limits, a suite of next-generation telescopes is being designed specifically for exoplanet studies, promising to expand our capabilities and understanding in the future.

# 1. The Diverse Range of Exoplanets

Ever since the telescope showed that the planets were worlds like our own and not just points of light in the sky, people have wondered whether there could be life on them as there is on Earth. Today that wonder is extended to exoplanets—planets orbiting stars outside our Solar System—and the search for signs of life on exoplanets is a key motivator in exoplanet research. Since the discovery of the first exoplanet around a Sun-like star in 1995, over 5,000 exoplanets have been identified, showcasing an astonishing variety in size, mass, and temperatures (i.e., orbits) (Figure 1). This diversity challenges our preconceived notions of planet formation and characteristics, and ultimately is expected to extend to a diversity of habitable conditions.

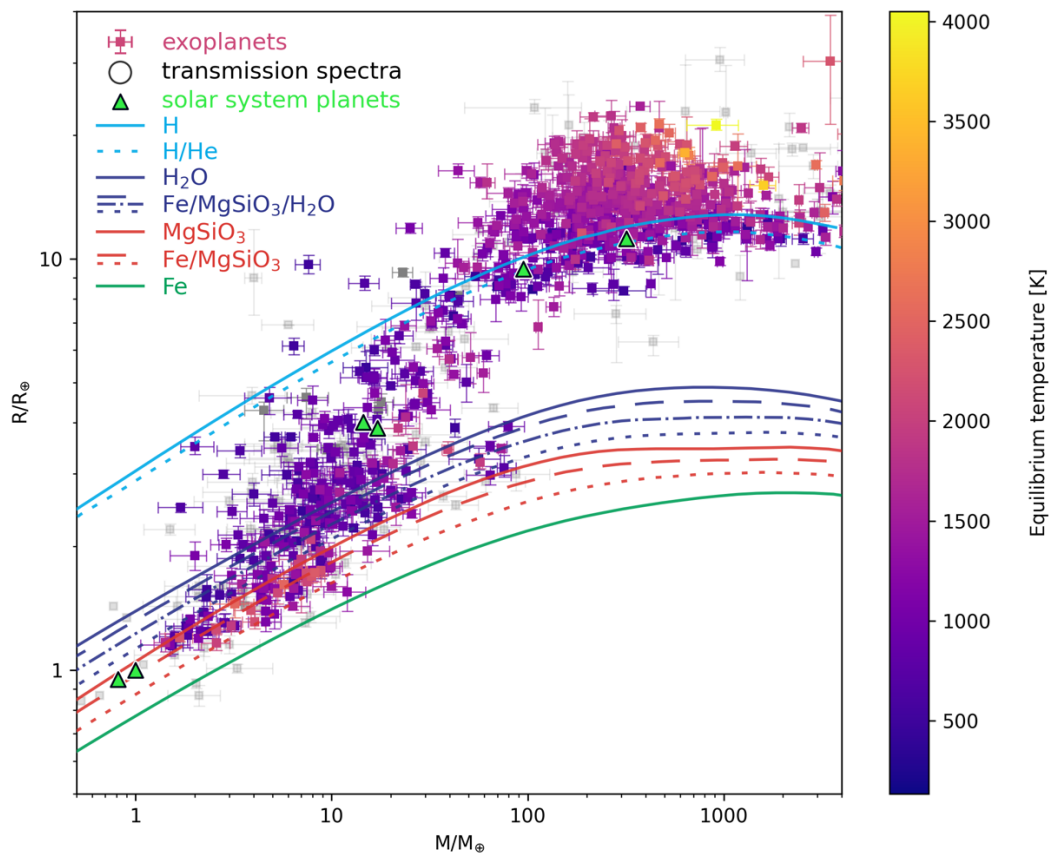

**Figure 1.** The diversity of exoplanets. Exoplanets with measured masses (x axis) and radii (y axis) are shown as points colored by their equilibrium temperature. Colored points are those with uncertainties less than 30% and grey are planets with uncertainties between 30% and 50%. The curves are illustrative models representing planetary compositions; if a planet aligns with a curve, its interior may match that composition. Although only a subset of the thousands of detected exoplanets have measured masses and radii, those shown here illustrate the theme of exoplanet diversity as captured by wide range of masses, radii, and orbits (via their temperatures). Data from the Exoplanet Archive https://exoplanetarchive.ipac.caltech.edu/ (Akeson et al., 2013). Models from (Seager et al., 2007). Credit: L. Herrington and S. Seager.

Exoplanets vary widely, from massive giants over ten times Jupiter's mass to small, rocky worlds smaller than Earth (Figure 1). This distribution includes sub-Neptune-sized (sub Neptunes) and super-Earth-sized planets—an enigmatic group with sizes between Earth and Neptune that defy easy classification due to their varied, ambiguous densities. Sub Neptunes lack any counterpart in our Solar System, yet surprisingly appear to be an extremely common planet type in our Galaxy. Some exoplanets orbit so closely to their host stars that they complete a revolution in less than an Earth day, while others take centuries, orbiting in the distant fringes of their planetary systems.

For decades, astronomers focused the search for life on the habitable zone—the region around a star where a planet's surface temperature could support liquid water, assuming a thin atmosphere and a rocky surface similar to Earth. Before knowing about the vast diversity of exoplanets, this definition often pointed towards an Earth analog (an Earth-sized planet in an Earth-like orbit about a Sun-sized star). An Earth-sized planet is minuscule compared to the Sun: 10,000 times smaller in area, 300,000 times less massive, 10 million times fainter in thermal (infrared) emission, and 10 billion times fainter in reflected light. Currently, Earth analogs are out of reach with current observational facilities and thus remain targets for future telescopes (Section 4).

While Earth analogs are currently beyond the reach of existing observatories, we are fortunate that exoplanet diversity allows us to focus on more detectable targets. The focus for habitable worlds this decade remains on the habitable zone, particularly around small M dwarf stars (Tarter et al., 2007). These stars, a tenth to half the size of our sun, simplify the detection and analysis of orbiting planets and their atmospheres as compared to Sun-sized stars and their planets (Nutzman and Charbonneau, 2008) (Figure 2). Despite initial skepticism, the potential for observation has encouraged a more favorable reassessment.

The planetary environments of M dwarf stars, however, differ significantly from those around Sun-like stars. Their planets, often tidally locked due to their close proximity to the host star, face a sharp divide between perpetual day and night and are also subject to strong flares and intense stellar radiation from magnetically active M dwarf stars, far stronger than any body in our Solar System.

Furthermore, during their tumultuous early phases, marked by intense heating and strong stellar winds, these planets are subjected to harsh radiation for hundreds of millions of years (Luger and Barnes, 2015; Shields et al., 2016). This brings into question whether these planets retain atmospheres capable of supporting life or if any original surface water has been irreversibly lost.

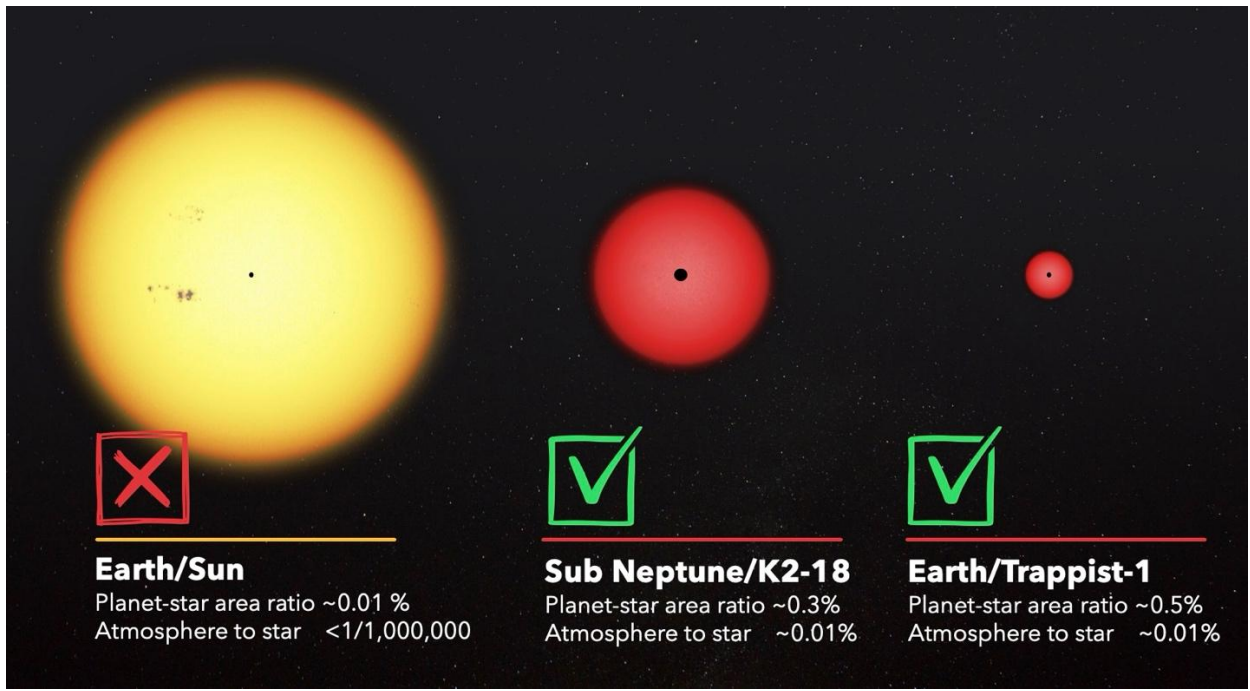

**Figure 2.** Red dwarf stars are more favorable for transiting planet atmosphere characterization. The main way we observe exoplanet atmospheres today is via transiting planets. The planet transit signal is the planet-to-star area ratio. The atmosphere signal is approximately the area of the planet atmosphere annulus to the background star. Both signals are 10 to 100 times larger (middle and right for two known planet systems; artist's conception) for planets transiting M-dwarf stars compared to an Earth transiting a Sun (left; real solar image, credit NASA).

Despite their observational advantages, M dwarfs present significant challenges for observations of transiting exoplanet atmospheres. A key issue is distinguishing planetary atmospheric signatures from the effects of stellar magnetic activity, such as star spots, which can heavily distort or mimic atmospheric signals (Figure 3). The challenges and potential mitigation strategies of M dwarf host star activity effects on small transiting exoplanet atmospheres are summarized in a report from NASA's Exoplanet Exploration Program Study Analysis Group 21 (SAG21; (Rackham et al., 2023) and an intensive focus of current astronomers' efforts.

Sub Neptune-size exoplanets transiting M dwarf stars present even more observationally favorable targets for transmission spectroscopy within the habitable zone than Earth-size exoplanets. Sub Neptunes are not only larger than Earth but have $H_2$-dominated atmospheres making them lighter (i.e., more puffy) and more accessible via transmission spectroscopy. However, the internal composition of sub Neptunes is not well understood; they could be water worlds, scaled-down versions of Neptune, or have mixed hydrogen envelopes overlying a magma ocean.

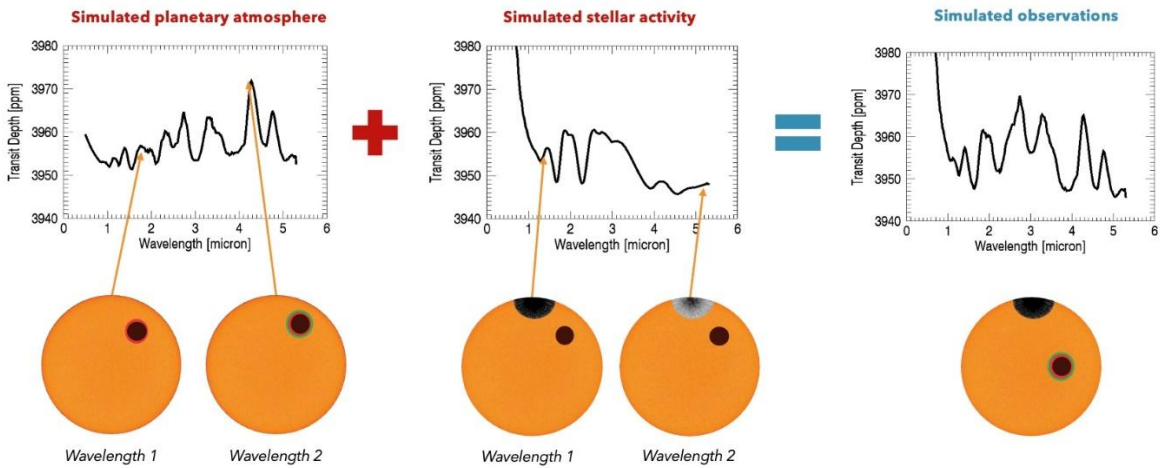

**Figure 3.** Illustration of star spot contamination contribution to a transit transmission spectrum. Left: a planet atmosphere changes the size of a planet as a function of wavelength, depending on the absorption strength and wavelength of atmospheric gases. This changes the planet+atmosphere to star area ratio—i.e., the transit depth—and creates what is known as the planet atmosphere transmission spectrum. Middle: assuming an atmosphereless planet for illustration, a star spot also changes the planet-to-star area ratio as a function of wavelength. This is because a star spot spectrum changes with wavelength and thus changes the apparent star area as a function of wavelength. Right: the contamination from star spots intermixes with the planet atmosphere signal; disentangling the two is a major current endeavor because M dwarf stars tend to be magnetically active with many evolving spots. Credit: S. Seager and A. Shapiro for the REVEAL Team.

Building on our exploration of diverse exoplanet environments and observational challenges, the recently operational James Webb Space Telescope (JWST; (Gardner et al., 2006)) is now actively pushing the boundaries of atmosphere detection. While JWST excels at analyzing the large atmospheres of hot giant planets, it is also extending its reach to smaller planets orbiting M dwarf stars. Despite the atmospheric signals being faint and at the limits of JWST's capabilities (Seager et al., 2025), ongoing efforts are effectively targeting several habitable-zone planets. Within the habitable zone Earth-size and sub Neptune size there is surely a broad spectrum of habitability potentials, shaped by diverse atmospheres and the absence of traditional solid surfaces—conditions we explore in this article.

## 2. Planet Atmosphere Types are Not Limiting for the Persistence of Life

Studies via computer simulations of habitable exoplanets and potential biosignature gases often assume atmospheres primarily composed of nitrogen ($N_2$)—as with Earth—and carbon dioxide ($CO_2$), similar to Mars and Venus. These models also account for various other gases, such as $O_2$ on modern Earth or $CH_4$ on early Earth, though these are typically present in only trace amounts, akin to $O_2$ on Mars and Venus. However, the presence of terrestrial life should not constrain our models too narrowly. This section explores the implications of dominant gas compositions in planetary atmospheres, focusing on the continuous viability of surface life on exoplanets, rather than subsurface life or the origins of life.

### *2.1 $H_2$-Dominated Atmospheres*

Rocky planets with $H_2$-dominated atmospheres are the most favorable for atmosphere observation via transit transmission spectra because the low density of $H_2$-gas (compared to, e.g., $N_2$ or $CO_2$) leads to an expansive atmosphere.

We do not know, however, if rocky exoplanets can maintain an $H_2$-dominated atmosphere, because under habitable surface temperatures the atmosphere is such that hydrogen will escape the planet to space. There is a consensus that terrestrial planets are formed with a significant amount of atmospheric $H_2$, including Earth (Young et al., 2023). Perhaps a massive terrestrial planet could maintain an $H_2$ atmosphere if it is continually replenished from outgassing of reduced interior reservoirs, or left over from a massive $H_2$ and He primordial envelope (see (Seager et al., 2020) and references therein, as well as (Chachan and Stevenson, 2018) and (Mulders et al., 2019)). Certainly, sub Neptunes have maintained an envelope dominated by $H_2$ or $H_2$-He mixtures.

$H_2$ is harmless to terrestrial life (i.e., not toxic) in either small or large quantities because it more or less behaves as an inert gas. An atmosphere composed entirely of $H_2$ would be rapidly fatal for terrestrial animals, but only because it contained no $O_2$, not because the hydrogen itself is toxic. We have demonstrated that single-celled microorganisms that normally do not inhabit $H_2$-dominated environments can survive and grow in a 100% $H_2$, entirely oxygen-free atmosphere (Figure 4; (Seager et al., 2020). We tested a model prokaryote (*Escherichia coli*) and a model eukaryote (yeast, *Saccharomyces cerevisiae*). Both are facultative anaerobes that are able to live in environments with and without $O_2$. Both *E. coli* and yeast grew at slightly lower rates in the $H_2$ environment than in air, as is expected given the lower efficiency of anaerobic metabolism based on either anaerobic respiration or fermentation compared to an aerobic energy capture. For yeast the lower growth rates are also because $O_2$ is a crucial substrate in the biosynthesis of many biochemicals essential to some eukaryotes. Aside from our study, and a recent report on the long-term growth and survivability of *E. coli* in 100% $H_2$ atmosphere (Kuzucan et al., 2025), it is worth noting that methanogens and acetogens are routinely grown in 80% $H_2$ and 20% $CO_2$ (Balch et al., 1979; Peters et al., 1998), where $H_2$ is a crucial reductant. Additionally, $H_2$ plays a central role as a reductant in many other anoxic organisms.

In the same work Seager et al. showed that *E. coli* and yeast can grow in pure He atmospheres, a somewhat contrived scenario that may exist in cases where H has escaped leaving an He atmosphere remaining (Hu et al., 2015).

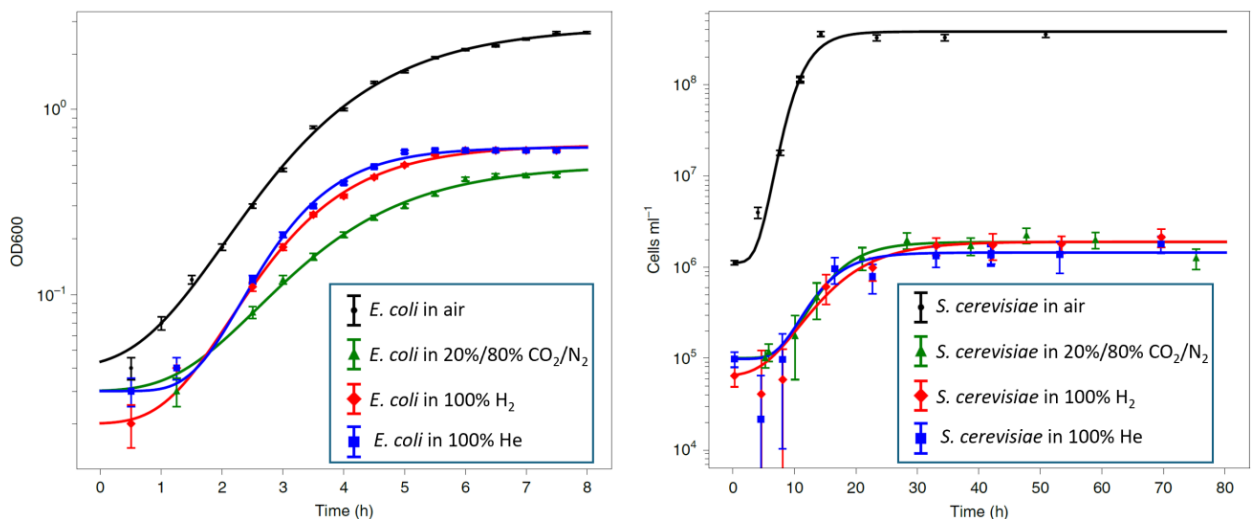

**Figure 4.** Growth curves of *E. coli* bacteria (left panel) and *S. cerevisiae* yeast (right panel). Y-axis: cell concentration in the bacterial or yeast culture in units based on the optical density at 600 nm (OD600). X-axis: time in hours. The growth curves are color coded according to the legends in both panels. The high concentrations of $H_2$ (and other gases) is not a limiting factor for the persistence and growth of microbial life. Figure adapted and modified from (Seager et al., 2020).

## 2.2 CO-Dominated Atmospheres

Several theoretical studies show that high concentrations of CO could accumulate on rocky exoplanets orbiting M-dwarf stars (Schwieterman et al., 2019b, 2019a). In addition, a "CO runaway" effect has been identified via models for early Earth (Kasting, 2014; Kasting et al., 1984, 1983; Zahnle, 1986). Moreover, without the Cl catalytic cycle that stabilizes $CO_2$ on Venus, Venus' atmosphere should have a substantial CO component (DeMore and Yung, 1982).

Despite CO's well-known toxicity to mammals, a result of CO inhibiting oxidative energy metabolism, microbial life on Earth can easily adapt to very high levels of CO. While on Earth natural ecological niches, created by various microbial colonies, likely do not exceed ~600 ppm of CO (Robb and Techtmann, 2018) many microbial species are completely unaffected by 100% headspace of CO. For example, *Thermoanaerobacter thermohydrosulfuricus*, a naturally occurring strain of thermophilic bacteria (Alves et al., 2016) and *Thermoanaerobacter kivui* can live in 100% CO and use it as a sole source of energy and carbon (Weghoff and Müller, 2016). As an interesting aside, CO is not only consumed by microbes to be used in energy metabolism but is also produced by microbes (although the reasons for the microbial production of CO are unclear) (Colman et al., 2011). Despite its acute toxicity, mammals have evolved to handle CO and even to produce it, for example CO is produced in very small amounts naturally by the human body as a signaling molecule (Untereiner et al., 2012).

It has been postulated that CO would prevent complex life from developing, as on Earth CO is a deadly poison for complex $O_2$-dependent life (Schwieterman et al., 2019a), a hypothesis strongly criticized by (Howell, 2019). CO is a rare gas on Earth, almost never encountered naturally by life, so there is no evolutionary pressure for $O_2$-dependent organisms to evolve to tolerate high levels of CO.

We logically argue that any life in high concentration CO atmospheres would successfully adapt, and possibly even exploit CO for its own benefits. The analogous situation happened with $O_2$ after Earth's Great Oxidation Event. Life adapted to survive in high $O_2$ conditions, and later exploited $O_2$ as a main component of its energy metabolism[1].

Furthermore, the toxicity of CO is highly specific to vertebrate biology and should not be extrapolated to the entire biosphere of complex life on Earth (Howell, 2019). In more detail, CO is highly toxic to mammals because it binds to iron in heme in hemoglobin (Hb). Hb is the principal $O_2$ carrier in vertebrate blood. Hb can bind to both $O_2$ and CO, but upon binding, Hb releases CO more slowly than $O_2$. This competitive inhibition of $O_2$ binding is why CO is a highly poisonous gas (Coburn, 1979). However, it has been known for more than 100 years that CO does not bind to all hemoglobins equally efficiently (Krogh, 1910). Moreover, not all animals use Hb as an $O_2$ carrier in their blood. Other $O_2$ carriers contain a variety of respiratory pigments, e.g. in hemocyanin (blue), hemerythrin (violet), erythrocruorin (red), chlorocruorin (green) and their affinity to CO varies between each carrier and species. For example, hemocyanin binds CO with lower affinity than Hb. This affinity also varies between different species, with horseshoe crab (*Limulus sp.*) hemocyanin having 100-fold lower affinity to CO than to $O_2$ (Ellerton et al., 1983; Van Holde and Miller, 1982). Likewise, the limited available data on hemerythrin resistance to CO suggests that hemerythrin from peanut worms (*Sipuncula*) has negligible binding to CO (Klotz and Klotz, 1955; Mot et al., 2010).

---

[1] $O_2$ may be essential for complex life due to its high-energy yield, but whether CO could similarly support energy metabolism remains uncertain. While complex life may not strictly require $O_2$ at sufficient pressure, once oxygenic photosynthesis evolves, life with high energy demands is likely to benefit from and become reliant on $O_2$.

It is unknown if CO is toxic to animals that have no blood pigment whatsoever, such as to Antarctic fish, e.g. *Chiondraco rastrospinosus*, that have no erythrocytes or Hb (Ruud, 1954; Sidell and O'Brien, 2006), but lack of Hb would suggest that they have a high tolerance to CO.

Moreover, while there have been few direct studies on the toxicity of CO to organisms other than mammals, the available data support the highly variable and species-dependent effect of CO on animal physiology. For example, the deep sea Toadfish (*Opsanus tau*) survive indefinitely in 60% CO (Wittenberg and Wittenberg, 1961) and freshwater goldfish (*Carassius auratus*) can survive exposure to 80% CO for at least two days at 5 °C with no obvious adverse effects (Anthony, 1961). *Xenopus laevis* frogs can withstand a nearly complete elimination of their Hb due to the toxic effect of CO, and subsequently recover without any apparent ill effects (De Graaf, 1957). Experiments on insects, like and common fruit fly (*Drosophila melanogaster*), show that the flies can be exposed to 95% CO for up to 6 hours without any detrimental effects (Clark, 1958).

The examples above clearly illustrate that the effect of CO on animal physiology varies and that acute toxicity of CO gas to mammals cannot be generalized across all animals, not to mention across all complex life.

## 2.3 $CO_2$-Dominated Atmospheres

Some rocky exoplanets are expected to have $CO_2$-dominated atmospheres resulting from carbon outgassing from accreted planetesimals (e.g., (Elkins-Tanton and Seager, 2008)), $CO_2$ outgassing from magma oceans (e.g., (Lammer et al., 2018)), and thermochemical equilibrium (e.g., (Schaefer and Fegley Jr., 2011)). In our Solar System, both Mars and Venus exhibit $CO_2$-dominated atmospheres, containing 95% and 96.5% $CO_2$ by volume, respectively. Unlike Mars and Venus, most of Earth's initial $CO_2$ inventory is locked in carbonate rocks, facilitated by $CO_2$'s dissolution in the ocean.

High $CO_2$-levels themselves are not universally toxic to life on Earth, and likely are also not universally detrimental to life elsewhere. The effects of high $CO_2$ levels on survival and growth of living organisms are complex and highly dependent on the species and the growth temperature (see e.g., the early work of (Coyne, 1933)). Dissolved $CO_2$ makes a medium more acidic, which often results in slower growth (Kuzucan et al., 2025). For example, elevated $CO_2$ concentrations have long been known to inhibit bacterial food spoilage and inhibit growth of several bacterial species (Dixon and Kell, 1989; Eyles et al., 1993; Gill and Tan, 1979; Koizumi et al., 1991; Leisner et al., 2007; Martin et al., 2003).

The inhibitory effect of high concentrations of $CO_2$, however, is not universal. Microbial species, including cyanobacteria, are capable of growth in > 95% by volume $CO_2$ atmospheres both at atmospheric pressures similar to Earth's as well as in a simulated low pressure Martian environment (Schuerger and Nicholson, 2016; Thomas et al., 2005). Adaptation of more complex organisms, like microalga *Chlorella vulgaris* (Likai et al., 2025), or lichens to simulated Martian conditions is also possible (de Vera et al., 2014). Microbial growth under supercritical $CO_2$ has also been reported (Peet et al., 2015). Most mammals are poisoned by $CO_2$ concentrations above 10%, the concentration varying widely with species and exposure time, and 5% $CO_2$ causes upper respiratory pain and increased heart and breathing rates (Permentier et al., 2017). However, this is a specific physiological adaptation to their normal, low-$CO_2$ environment, as illustrated by the burrow-dwelling naked mole rat finding a 5% $CO_2$ atmosphere entirely benign (Amoroso et al., 2023). Therefore, there is no reason to suppose that a $CO_2$-dominated environment is incompatible with complex life.

## *2.4 $O_2$-Dominated Atmospheres*

Massive $O_2$ atmospheres (with partial pressures $pO_2 \geq 10$ bar) are postulated to exist via abiotic accumulation on M-dwarf-star-hosted exoplanets. The main proposed mechanism is XUV-driven photodissociation and escape during the extended pre-main sequence runaway greenhouse phase (e.g., (Luger and Barnes, 2015)). M-dwarf stars experience an extended pre-main sequence phase that can last up to ~1 Gyr, during which both their bolometric luminosity and their XUV output are significantly enhanced. This means that planets which inhabit the liquid water habitable zone during the main sequence lifetime of the star will be interior to it during the pre-main sequence phase of the star. Such worlds are predicted to be in a runaway greenhouse state, in which the enhanced instellation evaporates the oceans and moistens the upper atmosphere.

In the upper atmosphere, the $H_2O$ is vulnerable to photodissociation to H and O; the H escapes, dragging some, but not all, of the O with it; the residual O can accumulate in the atmosphere as $O_2$. In the limit of no O escape and no uptake of $O_2$ by the underlying planet, (Luger and Barnes, 2015) estimate the buildup of 240 bar of $O_2$ per terrestrial-ocean-equivalent of water lost. Whether this limit is achieved is a matter of debate, and of the details of the planetary context (Meadows et al., 2018; Schaefer et al., 2016; Tian, 2015; Wordsworth et al., 2018).

If the runaway greenhouse co-exists with a magma ocean, then the magma ocean could scrub the photolytic $O_2$ from the atmosphere if the melt were reducing, and the magma overturn timescale short, (Gillmann et al., 2009; Luger and Barnes, 2015; Tian, 2015). For the specific case of GJ 1132b, (Schaefer et al., 2016) argue that most of the photolytic O should be lost to escape, and a significant part of the remainder of O would be lost to oxidation of the mantle mediated by the magma ocean; accumulation of substantial $pO_2$ requires high water bulk inventories, of ≥5% $H_2O$ by mass. (Wordsworth et al., 2018) explore abiotic atmospheric $O_2$ buildup more generally, and argue that such buildup is most likely for planets with (1) close-to-host-star orbits and (2) low mantle FeO and $H_2O$ inventories. We do not yet know if exoplanets with an abiotically generated $O_2$-dominated atmosphere exist but the possibility remains that they might exist on desiccated M dwarf terrestrial exoplanets (Gao et al., 2015). Many of the proposed abiotic mechanisms may lead to the buildup of only modest $O_2$ inventories (Reinhard et al., 2017; Wordsworth and Pierrehumbert, 2014; Wordsworth et al., 2018).

From a biosignature gas perspective, massive $O_2$ atmospheres are only proposed to be produced abiotically, and can be identified by the presence of $O_2$-$O_2$ dimers (Meadows et al., 2018; Schwieterman et al., 2016).

Some prokaryotes can survive 100% $O_2$ atmospheres at 1 atm pressure, such as *E. coli*, *Streptococcus* (*Enterococcus*) *faecalis*, *Bacillus subtilis*, and *S. cerevisiae*. The same species still survive, but have drastically, yet reversible, inhibited growth at 100% $O_2$) atmospheres above 1 atm pressure (Baez and Shiloach, 2014).

Unlike prokaryotes, eukaryotes are more sensitive to atmospheric oxygen concentration above 40% (as measured at 1 atm) (Lin and Miller, 1992). For example, cultures of human myeloid leukemia U-937 cells exposed to 50% of $O_2$ show strong growth inhibition compared with cells exposed to 21% $O_2$ (Scatena et al., 2004). We note however that most of the cells in the mammalian body experience substantially less than atmospheric $O_2$ concentration, so even exposing cells to 21% is equivalent to hyperoxia. Mammals, including humans, can survive over 1 bar of oxygen for short periods (for example, when SCUBA diving), but suffer toxic effects from breathing >0.3 bar for extended periods.

The adaptations needed for life to survive and thrive in very high $O_2$ atmospheres are simple to describe, although they may be difficult to implement. Very high levels of $O_2$ (> 1 atm of 100% for prokaryotes, and above 40% for eukaryotic cells) is tied to the generation of reactive oxygen species which at harmful levels can overwhelm the antioxidant defenses and repair systems of cells. In principle, life would only need to produce a greater concentration of reactive oxygen species scavengers to tolerate a higher $O_2$ concentration. Life already does this in hyperoxic conditions. In practice there may be a limit to how much of life's metabolism it can divert to protecting itself from oxygen.

High $O_2$ concentrations can also limit land life through the mechanism of fire risk. The chance of ignition and spread of fire increases dramatically with increasing oxygen partial pressure (Scott and Glasspool, 2006), and even relatively inert materials can burn extremely rapidly in a 1 bar oxygen atmosphere, as illustrated in the Apollo 1 fire. The limit this places on the abundance of atmospheric oxygen is unclear (Vitali et al., 2022). However, the feedback between land plants generating oxygen and oxygen-driven fire limiting those plants has been suggested as a limit for the oxygen concentration in the Earth's atmosphere to around 30% in the Carbonifeous (Mills et al., 2023), when fires were common (Scott, 2024).

| $N_2$-dominated atmospheres | $CO_2$-dominated atmospheres |
|---|---|
| $N_2$ is a dominant gas in Earth's atmosphere for the entirety of its geological history. $N_2$-dominated atmospheres are not only compatible with life's continuous persistence but also allow for its origin. $N_2$ is a chemically inert gas. | The inhibitory effect of high concentrations of $CO_2$ is not universal. Cyanobacteria can grow in > 95% $CO_2$ atmospheres, both at Earth's atm. pressures and in a simulated Martian low pressure environments. Adaptation of more complex organisms, like lichens, to simulated Martian conditions is also possible. Microbial growth under supercritical $CO_2$ has also been reported. |
| **CO-dominated atmospheres** | $H_2$**-dominated atmospheres** |
| Many microbial species are completely unaffected by 100% CO atmospheres. For example, *T. thermohydrosulfuricus* and *T. kivui* can live in 100% CO and use it as a sole source of energy and carbon. | $H_2$ is not toxic. Life can routinely survive and grow in a 100% $H_2$ atmosphere. For example, *E. coli* and *S. cerevisiae*, and many other species. |
| $O_2$**-dominated atmospheres** | **Any other atmospheres?** |
| Many species can survive and reproduce in 100% $O_2$ atmospheres at 1 atm pressure, such as *E. coli, S. faecalis, B. subtilis*, and *S. cerevisiae*. | Life can easily adapt to atmospheres with artificial gas composition. Atmospheric composition is not a limiting factor for the persistence of life. |

**Table 1.** Summary of Earth life's tolerance to different gases, supporting the concept that life can thrive in a wide variety of planetary atmosphere composition. See text for references.

## *2.5 $N_2$-Dominated Atmospheres*

For completeness we briefly acknowledge $N_2$-dominated atmospheres (since Earth's is). Earth has retained its $N_2$-dominated atmosphere for the entirety of its geological history. This includes the Archean eon when the atmosphere of Earth had a reducing character, possibly containing significant amounts of $CH_4$, (Catling and Zahnle, 2020), as well as the oxidized atmosphere after the Great Oxygenation Event. For this entire time Earth was not only habitable but also, during the Archean eon, had conditions conducive to the origin of life. We can therefore conclude that the $N_2$-dominated atmospheres are not only compatible with life's continuous persistence but also allow for its origin. We also note that $N_2$ is the most chemically inert gas other than the Group 8 'noble' gases present in any known planetary atmosphere, and so is unlikely to cause a toxicity hazard to any biochemistry.

## *2.6 Other Gases*

Life can easily survive and thrive in conditions where gases other than those discussed above are minor components of the environment. Many methanotrophs can routinely grow in 20% CH$_4$ headspace conditions (e.g., (Danilova et al., 2016; Farhan Ul Haque et al., 2018; Vekeman et al., 2016)) and *E. coli* cultures have been reported to grow in 80% CH$_4$ atmosphere (Kuzucan et al., 2025). Microbial growth in elevated concentration of H$_2$S (0.4% in headspace) and methane (20%) has also been reported (Jiang et al., 2023). While exoplanet atmospheres where CH$_4$, H$_2$S and other minor constituents are dominant gases are unlikely to exist, the fact that life can easily adapt to such artificial conditions further supports the conclusion that atmospheric composition is not a limiting factor for the persistence of life (Table 1).

## 3. Further Beyond Terracentricity

In the spirit of expanding our prospects for life detection, we further consider the diversity of habitable worlds beyond atmosphere types and to non-water solvents and planets without temperate surfaces.

### *3.1 Beyond Water as a Solvent for Life*

Planets and moons with liquids other than water exist in the Solar System. Liquid methane and ethane fills the seas, lakes, and rivers on Saturn's moon Titan (Stofan et al., 2007). Liquid droplets of concentrated sulfuric acid are the main component of Venus' clouds (Titov et al., 2018). In contrast to Titan's liquid hydrocarbons (McKay, 2016; McKay and Smith, 2005), concentrated sulfuric acid as a planetary liquid has received the least attention from the astrobiological point of view (for a recent overview of alternative solvents for life see (Bains et al., 2024a)). This is despite the fact that concentrated sulfuric acid as a planetary solvent is the dominant liquid on Venus and could be one of the most common liquids in the Galaxy (Ballesteros et al., 2019), although the stability of concentrated sulfuric acid as a surface liquid is unknown.

Earth does not have an environment that is analogous to Venus sulfuric acid clouds. No life on Earth would be able to survive in Venus clouds as the droplets are many orders and magnitude more acidic than even the most acidic environments inhabited by Earth's extremophiles (reviewed in (Seager et al., 2021a)). Despite these challenges, recent studies suggest that concentrated sulfuric acid is not hostile to complex organic chemistry. It can sustain diverse organic chemicals dissolved in it, including amino acids (Seager et al., 2024a), dipeptides (Petkowski et al., 2025, 2024), nucleic acid bases (Seager et al., 2024b, 2023) and lipids, capable of forming complex vesicle-like structures in concentrated sulfuric acid (Duzdevich et al., 2025). Rich organic chemistry in concentrated sulfuric acid and the possible presence of organic chemistry in the sulfuric acid clouds of Venus (Spacek et al., 2024) exemplify how planets with environments very different than Earth can still be astrobiological targets for the search for life, albeit life with biochemistry entirely distinct from Earth's. Combining laboratory studies on organic chemistry in concentrated sulfuric acid, with remote observations of Venus and space missions focusing on the chemical composition of the clouds could redefine planetary habitability.

We now turn to the prospects of detectability and propose possible strategies to detect worlds where sulfuric acid could be a dominant liquid. Concentrated sulfuric acid can only accumulate on a planet that is substantially depleted of water. Such planets could be common around mature M dwarfs. The activity of the young M dwarf stars could bake their planets dry within the span of a few tens of millions to hundreds of millions of years (Luger and Barnes, 2015; Shields et al., 2016). Subsequent volcanic emission of $SO_2$ and $H_2SO_4$ would then generate sulfuric acid which could condense into surface liquid or cloud aerosols.

Sulfuric acid is not volatile and as such is difficult to directly detect remotely. To confirm that a candidate planet could host conditions compatible with either surface liquid sulfuric acid or concentrated sulfuric acid condensed into cloud droplets we would have to first establish that the planet is sufficiently devoid of water (beyond the lack of atmospheric water vapor). One such gaseous indicator of a dry planet is silicon tetrafluoride ($SiF_4$). On Earth, a planet with abundant water, the main fluorine containing volcanic gas is HF. $SiF_4$ is still present in volcanic gases, but it is only occasionally an abundant product of volcanic eruptions (Francis et al., 1996; Mori et al., 2002). In contrast, $SiF_4$ is expected to be a major F-containing volcanic gas on a planet that is H-depleted. HF, $H_2O$, HCl and other H-containing gases will still be produced by volcanoes, but the $SiF_4$ will be a significant volcanic product. Simultaneous observations of HF, $H_2O$, HCl and $SiF_4$ should confirm if a candidate sulfuric acid world is H-depleted. The IR spectrum of $SiF_4$ is known (Francis et al., 1996). $SiF_4$ has a dominant sharp feature around 9.7 μm, within JWST's MIRI-LRS spectral band and in a wavelength region of 9-12 μm with no expected major dominant atmosphere gases with strong spectral features, although many other trace gases do have features in this window.

In summary, research on alternative solvents is important not only in expanding our understanding of habitable environments but also in broadening the scope of potential targets in the search for extraterrestrial life.

### *3.2. Is a Rocky Surface Needed for Life?*
The most observationally favorable targets for JWST transmission spectroscopy within the habitable zone are sub-Neptune-size exoplanets, which are larger than Earth and have $H_2$-dominated envelopes, both factors making their atmospheres more accessible for study. The internal composition of sub Neptunes remains unclear—they could be water worlds, scaled-down versions of Neptune, or possess mixed hydrogen envelopes overlying a magma ocean.

Regardless, it is generally accepted that they do not have a rocky temperate surface. Moreover, the standard definition of the habitable zone, implying temperatures suitable for liquid water under a thin atmosphere in thermal equilibrium with stellar radiation, fails to consider the potent greenhouse effect of $H_2$-rich envelopes. Nevertheless, it is speculated that sub Neptunes may contain deep global water oceans or water clouds potentially capable of supporting life, prompting us to question whether a rocky surface is indeed essential for habitability.

Life in the water clouds is a conceivable scenario for sub Neptune-sized exoplanets, where organisms would persist well above the lethally hot depths, similar to the models proposed for potential life in Venus' clouds (Seager et al., 2021a). However, the existence of a fully aerial biosphere remains uncertain. While Earth's clouds support diverse microbial life, it's unclear if they provide more than just a means of transportation, rather than serving as permanent habitats (Lappan et al., 2024). For clouds to be continuously habitable they would need to maintain patches of permanent cloud cover, unlike Earth's transient cloud formations. For an overview of the challenges that face the hypothetical strictly aerial biosphere see (Bains et al., 2024b; Seager et al., 2021a), in the context of Venus clouds, and (Seager et al., 2021b), in the context of sub Neptunes.

In brief, sub Neptunes may host permanent patches of water clouds (Charnay et al., 2021) capable of supporting an aerial biosphere, despite the challenges to life's persistence in such settings (Seager et al., 2021b). The chemical reactions that could lead to the origin of life in these cloud environments remain a distinct topic with several hypotheses proposed (e.g. (Woese, 1979)). For example, water micro-droplets have been postulated as a potential environment for prebiotic formation of sugar phosphates and uridine ribonucleoside, precursors for nucleic acids (Nam et al., 2017), amino acid glycine or uracil (Meng et al., 2025). Life's emergence and sustainability are theorized to require access to minerals from a planet's surface. These minerals provide essential metal ions used by all known life forms in their structural and metabolic processes (Hoehler et al., 2020). Additionally, geological activity is crucial for the recycling of nutrients and biogenic elements.

The habitability argument in favor of a temperate rocky surface is that life requires inorganic nutrients such as metal ions that ultimately come from minerals.

A second scenario is that sub Neptunes might host life within global deep water oceans, a scenario grounded by models proposing these planets as water worlds or "Hycean Worlds"—featuring hydrogen envelopes atop global water oceans ((Madhusudhan et al., 2021; Rogers and Seager, 2010)). However, many (or most) known sub Neptunes possess thick hydrogen/helium envelopes that create greenhouse effects, rendering the ocean beneath potentially too hot and even supercritical—conditions hostile to life, where water acts more like a biocide by breaking down organic materials into simple gases, leaving behind only a soot-like residue (Chakinala et al., 2010; Matsumura et al., 2005; Yoshida et al., 2004).

The viability of life in such vast oceans faces the challenge of dilution; essential nutrients and biochemicals remain too dispersed to sustain life without mechanisms to concentrate them. Earth-like hydrothermal vent systems, which recycle and concentrate nutrients, demonstrate how subsurface geological activity might support life by creating micro-niches in ocean floor rocks where vital chemicals could cluster. However, a planet with its water ocean sandwiched between layers of high-pressure ice does not have direct contact with the planet's rocky mantle and as a result lacks these interactions. If the exchange of materials between the rock and the ocean through the ice layer is not efficient then such an ocean is an unlikely host for a thriving biosphere (Hernandez et al., 2022; Journaux, 2022).

## 4. Observational Prospects

Mapping planetary characteristics to observables is a challenge, because there is usually no one-one correspondence. While grappling with the possible diversity of habitable worlds and robust biosignature gases, we face the reality that telescopes cannot provide all the answers. Instead, the limits of engineering and the inherent characteristics of stars shape what we can observe. Here we summarize the next-generation telescopes that are being designed to study exoplanet atmospheres, starting with the current JWST.

### *4.1 JWST*

The JWST is an infrared telescope with an effective aperture of 6.5 meters, currently orbiting 1 million miles from Earth at the Earth-Sun Lagrange 2 point. JWST's relevant wavelength range spans from 0.7 to 16 microns. JWST is equipped to observe the atmospheres of solar system-aged temperate planets through transit transmission spectroscopy, where the planet crosses in front of its host star from our perspective, imprinting its atmospheric signature on the starlight. While this method has proven most successful for hot giant planets with expansive atmospheres, approximately a dozen rocky exoplanets within their stars' habitable zones could potentially be studied using JWST, given the current noise floor levels. These rocky planets are limited to small M dwarf host stars (Section 1 and Figure 2).

Currently, JWST transit transmission spectra observations for M dwarf star hosts face significant challenges due to stellar contamination. Red dwarf stars are magnetically active, and frequently exhibit starspots which evolve over time, leading to an inhomogeneous stellar background. The wavelength-dependent spectrum of starspots differs from quiet star regions, enough to severely contaminating the tiny signal in transmission spectra of small temperature planet atmospheres (Figure 3). This active star contamination issue is a primary focus of ongoing research (Rackham et al., 2023). To expand the roster of potential habitable planet candidates, researchers are also concentrating on sub Neptune-sized exoplanets, many favorable targets oft which also orbit red dwarf stars.

## 4.2 E-ELT

The next major advancement in astronomy, following the JWST, is the construction of the Extremely Large Telescope (EELT, 39 m aperture diameter) (Gilmozzi and Spyromilio, 2007; Tamai and Spyromilio, 2014). The EELT can circumvent stellar contamination by directly imaging habitable-zone planets orbiting M dwarf stars, blocking out the starlight utilizing advanced coronagraph instrumentation and extreme adaptive optics. However, achieving the necessary planet-star contrast ratio of $10^7$ to $10^8$ presents significant challenges. The EELT's first-light instrument Mid infrared ELT Imager and Spectrograph (METIS (Brandl et al., 2016)) will feature infrared direct imaging capabilities (3 to 14 microns), a coronagraph, and extreme adaptive optics including low- and medium-resolution spectroscopy. Although METIS is optimized for large planet discovery through imaging, it may be able to access one to a few bright target stars for habitable-zone Earth-sized planets (Bowens et al., 2021). A second-generation instrument planned for the ELT, the ArmazoNes high Dispersion Echelle Spectrograph (ANDES), is designed for near-infrared imaging and spectroscopy (0.35 to 2.4 μm) to detect reflected light from rocky planets in the habitable zones of their stars. This capability could enable observations of up to about 100 nearby low-mass stars, primarily mid M dwarfs (Artigau et al., 2018; Guyon, 2018; Guyon et al., 2019).

In addition to EELT, the US large ground-based telescopes, the Thirty Meter Telescope (TMT, 30 m aperture diameter)(Sanders, 2013) and the Giant Magellan Telescope (GMT, 20 m aperture diameter)(Bernstein et al., 2014; Johns et al., 2012) are under development. Other than direct imaging, the large ground-based telescopes might be capable of a combination of high-dispersion, high-spectral resolution ($R$~100,000) spectroscopy with moderate high-contrast imaging to observe spectra of a few rocky planets orbiting Sun-like stars (Snellen et al., 2015). Note that not all IR wavelength regions are easily accessible from Earth's surface due to absorption from Earth's own atmospheric gases.

*4.3 Habitable Worlds Observatory*

The astronomy community aims to expand our scope from M dwarf to Sun-like stars, striving to discover an Earth twin—an Earth-sized planet in an Earth-like orbit about a Sun-like star, complete with a thin atmosphere, oceans, and continents. NASA's planned Habitable Worlds Observatory (HWO; https://www.greatobservatories.org/hwo), part of the NASA Great Observatories, is intended for a mid-2040s launch. This telescope, designed to directly image exoplanets around Sun-like stars for both discovery and atmospheric analysis, will feature a primary mirror approximately 6 meters in diameter. It will employ a coronagraph to eliminate starlight, enabling direct imaging of planets. The HWO will operate with low-resolution spectroscopy across visible wavelengths, from the blue (or UV) to the near-infrared.

*4.4 Other Telescope Concepts*

Several innovative telescope concepts are in development to discover and characterize Earth analog exoplanets. The space-based Large Interferometer for Exoplanets (LIFE; (Quanz et al., 2022)) would to operate within the 4 to 18.5 micron range, covering infrared windows rich in trace gas spectral features. LIFE plans to employ four elements, each with an aperture of 2 to 3.5 meters, along with a combiner spacecraft (Kammerer et al., 2022). Another concept, Starshade, involves a large, specially shaped screen, functioning as a starlight suppressor by formation flying tens of thousands of kilometers from a UV-visible wavelength space telescope, ensuring only planetary light is captured, which significantly enhances atmospheric characterization. Namely, because only planet light enters the telescope, the Starshade enables high throughput and simultaneous broad-band coverage (e.g., (Seager et al., 2015)). Starshade can work with any space telescope, providing the telescope is "starshade ready" with an acquisition camera, communications system to work with starshade, and appropriate spectrograph instrument. An additional concept, the Nautilus telescope project, aims to deploy a constellation of 8-meter, low-cost diffractive telescopes to study 1000 transiting Earth analogs, showcasing the ambition of future astronomical endeavors (Apai et al., 2022).

## 5. Summary

A triumph of astronomy has been the discovery of thousands of exoplanets, with a remarkable diversity far beyond planetary scientists' initial imagination. As we enter the era of atmospheric characterization, we can look to the variety of Earth life, which thrives under a broad range of atmospheric conditions. Although water is the sole solvent for life on Earth, biomolecules have shown stability in concentrated sulfuric acid and even in liquid methane and ethane, broadening the prospects for extraterrestrial life. However, not every aspect of diversity may support life; for instance, metal ions, essential for life, imply the necessity of surface contact or efficient meteoritic delivery to sustain an active biosphere. A new generation of telescopes equipped with advanced instrumentation is poised to enhance our understanding of small exoplanets, specifically by characterizing their atmospheres to assess habitability and detect potential biosignature gases.